\documentstyle[12pt,axodraw,color,euler,epsfig,times]{article}

\newcommand{\be}{\begin{equation}}
\newcommand{\ee}{\end{equation}}
\newcommand{\bea}{\begin{eqnarray}}
\newcommand{\eea}{\end{eqnarray}}

\newcommand{\bra}{\langle}
\newcommand{\ket}{\rangle}
\newcommand{\mycdot}{\!\cdot\!}
\newcommand{\ubar}{\overline{u}}
\newcommand{\tl}{\tilde{\lambda}}

\begin{document}

% --------------------------------------------------------------- shortcut declarations

\newcommand{\nc}[2]{\newcommand{#1}{#2}}

% --------------------------------------------------------------- header and contents
\begin{center}
\begin{bf}
\begin{Large}
Diagrammatic proof of the BCFW recursion relation for gluon amplitudes in QCD.
\end{Large}\\
\vspace*{\baselineskip}
Petros D. Draggiotis$^{(a,c)}$\footnote{{\tt pdrag@phys.uoa.gr}},
Ronald H.P. Kleiss$^{(b)}$\footnote{{\tt R.Kleiss@science.ru.nl}},
Achilleas Lazopoulos$^{(b)}$\footnote{{\tt A.Lazopoulos@science.ru.nl}} and
Costas G. Papadopoulos$^{(c)}$\footnote{{\tt costas.papadopoulos@cern.ch}}\\
\vspace*{\baselineskip}

$^{(a)}$University of Athens,
Physics Department, Nuclear \& Particle Physics Section, $GR-15771$, Athens,
Greece\\

$^{(b)}$IMAPP, Institute of Mathematics, Astrophysics and Particle Physics,
Radboud University, Nijmegen, the Netherlands\\

$^{(c)}$Institute of Nuclear Physics, NCSR "Demokritos", 15310, Athens, Greece
\end{bf}
\end{center}

%\tableofcontents

%--------------------------------------------------------------- Abstract
\abstract{
We present a proof of the Britto-Cachazo-Feng-Witten tree-level recursion relation for gluon amplitudes in QCD, based 
on a direct equivalence between BCFW decompositions and Feynman diagrams. We demonstrate that 
this equivalence can be made explicit when working in a convenient gauge. 
We exhibit that gauge invariance and the 
particular structure of Yang-Mills vertices guarantees the validity of the BCFW construction.
}

%--------------------------------------------------------------- Introduction
\section{Introduction}

During the last year much progress has been made in the understanding of analytical calculations
of dual amplitudes in perturbative Yang-Mills theories. Led by an observation of Witten \cite{Witten},
Britto, Cachazo and Feng (BCFW) have proposed a new recursion relation for tree amplitudes of gluons \cite{BCF} that naturally arrives at the simplest known expressions for some of those amplitudes in terms of Weyl - Van der Waerden spinor products. Explicit calculations have been performed using this technique \cite{qcd_calculations},\cite{fermions}, extensions to amplitudes involving particles from the electroweak sector \cite{ew_sector} have been pursued and a new approach to one loop amplitudes has been proposed \cite{one_loop} employing MHV vertices and unitarity arguments.

The BCFW
recursion relation features some remarkable characteristics, among which the on-shell analytic continuation
of selected off-shell propagators, the analytic continuation of two selected external momenta in the complex plane and
a decomposition of a color helicity amplitude into smaller helicity amplitudes with complex external momenta that doesn't appear to be in direct connection with the decomposition in Feynman diagrams. Moreover, the BCFW relation leads to the use of only three-point
(modified) vertices as building blocks of the theory, thus raising questions about the fundamental nature of the Yang-Mills four point vertex.

A proof of the BCFW relation was given  by Britto, Chachazo, Feng and Witten \cite{proof} which, though it made the connection between the analyticity properties of the color amplitude and the BCFW decomposition obvious, shed little light towards the way the latter relates to the usual Feynman diagram approach.

The simplicity of the final result for tree level gluonic amplitudes implies the existence of major cancellations between Feynman graphs. Since the BCFW formalism avoids those cancellations altogether it is advantageous to find a diagrammatic proof of the BCFW decomposition in order to understand these cancellations in the level of Feynman diagrams. A number of questions immediately arise. Is a diagrammatic proof  gauge dependent? If so, are there preferable gauges where simplifications take place? Which gauges are those? Is this decomposition exclusively working for pure YM theories, and if so, what is the key feature of the YM vertices that makes it work? Can we find a BCFW type recursion relation for scalar theories?

We intend to address these questions in the present paper as well as giving a generic diagrammatic proof of the BCFW decomposition. The paper is organized as follows: after clarifying some notation issues in section $2$,  we present, in section $3$ an analysis based on diagram enumeration that hints 
towards a connection between BCFW decompositions and Feynman graphs. In section $4$ we prove some important kinematical identities that will 
support the main parts of the diagrammatic proof (sections $6$ and $7$), whereas section $5$ describes the gauge in which we are working and the consequences of that choice. We 
conclude with some remarks on the generality of the proof and the role of gauge invariance in this analysis in section $8$.

%--------------------------------------------------------------- Notation
\section{Notation}
All amplitudes mentioned in this paper are the color-ordered amplitudes that result after the usual trace-based color decomposition \cite{color_decomposition}. We will conveniently omit all coupling constants and factors of $i$ as well as the color antenna of
all color-ordered amplitudes in what follows.

In non-vanishing tree-level color-ordered amplitudes, one can, without loss of generality, pick out a pair of adjacent legs with opposite
helicity and label the positive helicity one by $1$ and the negative helicity one by $n$. Unless otherwise specified we will use
this labeling throughout this paper.

The Dirac four-spinors used in this paper are the usual $u_{\pm}(p)$, $\ubar_{\pm}(p)$ with the important note that the relation
between $u(p)$ and $\ubar(p)$ is {\emph not} one of complex conjugation. 

The relation with Weyl - Van der Waerden spinors, when using the Weyl representation of $\gamma$-matrices, is 
\bea
u_+(p)=(0,\tilde{\lambda}_{\dot{b}}(p)\epsilon^{\dot{b}\dot{a}}) &\;\;& \ubar_+(p)=(\lambda_{b}(p)\epsilon^{ba},0)\\
u_-(p)=(\lambda_a(p),0)&\;\; &\ubar_-(p)=(0 ,\tilde{\lambda}_{\dot{a}}(p))
\eea

Four-momenta can be written in terms of Weyl spinors as 
\be
p_{\mu}=\frac{1}{2}\ubar_-(p)\gamma_{\mu}u_-(p)=\frac{1}{2}\tilde{\lambda}_{\dot{a}}(p)\overline{\sigma}_{\mu}^{\dot{a}a}\lambda_a(p)
\ee
and products of four-vectors can be performed using the identity
\be
\overline{\sigma}^{\mu,\dot{a}a}\overline{\sigma}_{\mu}^{\dot{b}b}=2\epsilon^{\dot{a}\dot{b}}\epsilon^{ab}
\ee
where $\overline{\sigma}^{\mu}=(\cal{I},-\vec{\sigma})$,  and similar identities for the $\sigma$'s.

Note that, since 
\be
\overline{\sigma}^{\mu,\dot{a}a}\sigma_{\mu,b\dot{b}}=2\delta^{\dot{a}}_{\dot{b}}\delta^a_b
\ee
we can write
\be
(p\mycdot\sigma)_{a\dot{a}}=p^{\mu}\sigma_{\mu,a\dot{a}}=\lambda_a(p)\tilde{\lambda}_{\dot{a}}(p)
\ee

Moreover, the usual conventions for spinor products\footnote{Note the minus sign in the $[ij]$ definition with respect to the definition in \cite{Dix01}.} will be adopted:
\bea
\bra ij \ket \equiv  \ubar_+(p_i)u_-(p_j)=\lambda_{a}(p_i)\lambda_b(p_j)\epsilon^{ab}\\
\left[ij\right] \equiv -\ubar_-(p_i)u_+(p_j)=\tilde{\lambda}_{\dot{a}}(p_i)\tilde{\lambda}_{\dot{b}}(p_j)\epsilon^{\dot{a}\dot{b}}
\eea

This results to 
\be
2 p\mycdot q = \bra pq\ket[pq]
\ee

A fairly general definition of the external gluons' polarization vectors\footnote{that corresponds to the light-like axial gauge} 
is (see \cite{Dix01},\cite{Kleiss01} for similar definitions.)
\bea
\epsilon_+^{\mu}(p)=\frac{\ubar_+(\overline{p})\gamma^{\mu}u_+(p)}{\sqrt{2}\ubar_+(\overline{p})u_-(p)}=
\frac{\tl_{\dot{a}}(p)\overline{\sigma}^{\mu,\dot{a}a}\lambda_a(\overline{p})}{\sqrt{2}\bra \overline{p} p\ket}\nonumber\\
\epsilon_-^{\mu}(p)=\frac{\ubar_-(\overline{p})\gamma^{\mu}u_-(p)}{\sqrt{2}\ubar_-(p)u_+(\overline{p})}=
-\frac{\tl_{\dot{a}}(\overline{p})\overline{\sigma}^{\mu,\dot{a}a}\lambda_a(p)}{\sqrt{2}[p\overline{p} ]}
\label{polarization_vectors_def}
\eea
where $\overline{p}$ is an auxiliary, null four-vector that can be chosen at will for every gluon as long as it is not 
parallel to $p$ itself.

Note that one could write
\bea
\epsilon^+_{a\dot{a}}(p)\equiv\epsilon^+_{\mu}(p)\sigma^{\mu}_{a\dot{a}}=\sqrt{2}\frac{\lambda_a(\overline{p})\tl_{\dot{a}}(p)}{\bra \overline{p}p\ket}\\
\epsilon^-_{a\dot{a}}(p)\equiv\epsilon^-_{\mu}(p)\sigma^{\mu}_{a\dot{a}}=-\sqrt{2}\frac{\lambda_a(p)\tl_{\dot{a}}(\overline{p})}{[ p\overline{p}]}
\eea
and perform vector products using 
\be
p_{a\dot{a}}q_{b\dot{b}}\epsilon^{ab}\epsilon^{\dot{a}\dot{b}}=\bra pq\ket[pq]=2 p\mycdot q
\ee
It should be noted that nothing of what follows depends on the precise conventions we adopt for 
the Van der Waerden or Dirac spinors. In fact the present analysis, that deals with pure gluonic amplitudes,
could be performed exclusively on the level of four-vectors.

The hat symbol over a function of momenta $\hat{f}(q)$ will denote the function $\hat{f}(q;z)$ where the argument of $f$
is analytically continued (shifted) by a  general four-vector $z\epsilon_{\mu}$,
with
\be
\epsilon_{\mu}\equiv \frac{1}{2}\ubar_-(p_1)\gamma_{\mu}u_-(p_n)=\frac{1}{2}\tilde{\lambda}_{\dot{a}}(p_1)\overline{\sigma}^{\dot{a}a}_{\mu}\lambda_a(p_n)
\label{e_definition}
\ee

This has the effect
\be
q\rightarrow\hat{q}=q+z\epsilon
\ee

We will omit the explicit reference to the $z$-dependence of $\hat{f}(p)$ since this is signified by the hat symbol.

%In what follows we use the expression `hatted diagram' to denote a diagram that is seen as the real restriction of the corresponding %complex function that results by extending some of its four-momenta arguments to the complex plane. Those arguments are shifted according %to the hat procedure described above, and, 
%furthermore, an expansion along the poles of the resulting function is taken.
%Each pole corresponds to a value of $z\rightarrow z_0$, for $z_0$ such that one of the diagram's propagators (the `cut' one) becomes %on-shell.

%--------------------------------------------------------------- Counting graphs
\section{Counting contributions}
\label{counting}

The BCFW recursive relation for pure gluon color ordered amplitudes
is given by

{\small
\bea &&{\cal A}(p_1^{\lambda_1},\ldots,p_n^{\lambda_n})= \\
&&\sum_{\lambda_X=\pm 1}\sum_{k=1}^{n-1} {\cal
A}(\hat{p}_1^{\lambda_1},\ldots,p_k^{\lambda_k},-\hat{p}_X^{-\lambda_X})
\frac{1}{P_{1\ldots k}^2} {\cal
A}(\hat{p}_X^{\lambda_{X}},p_{k+1}^{\lambda_{k+1}},\ldots,\hat{p}_n^{\lambda_n})
\nonumber\eea
}
where $p$ and $\lambda$ represent momentum and polarization. 
In this section we would
like to study the diagrammatic content of the BCFW equation. The most
appropriate way to do so, is of course to neglect any reference to
polarization and also drop the propagator factor, while keeping the
ordering of the momenta. In that case of course we count planar
graphs (see \cite{'tHooft:1998mh}) and it is obvious that the equation is not self-consistent.
In fact the form of the equation is given by
 \be
{\cal A}=\sum
\begin{picture}(130,30) (35,-12)
\SetScale{0.5}
\SetWidth{0.5} \SetColor{Black} \GOval(135,-15)(30,30)(0){0.411}
\Line(75,-15)(105,-15) \Line(165,-15)(195,-15)
\GOval(135,-15)(30,30)(0){0.411} \GOval(225,-15)(30,30)(0){0.411}
\Line(75,-15)(105,-15) \Line(255,-15)(285,-15)
\Text(60,-10)[lb]{\small{\Black{${\cal A}_I$}}}
\Text(105,-10)[lb]{\small{\Black{${\cal A}_{II}$}}}
\end{picture}
\ee

or in a more mathematical form

\be {\cal A^{BCF}}_{1\to n}=\sum_{k=1}^{n-1} {\cal A}_{1\to
k+1}{\cal A}_{1\to n-k+1 } \ee

The equation ${\cal A^{BCF}}_{1\to n}={\cal A}_{1\to n}$ cannot be
true because of the following reasons:
\begin{itemize}
\item
Firstly, contributions where no propagator line
exists between the $1$st and the $n$th particle, are not included in
the BCFW equation
\be
{\cal A}^0=
\begin{picture}(40,45) (20,-3)
\SetScale{0.5}
\SetWidth{0.5} \SetColor{Black} 
 \Line(45,0)(75,0)
\Line(75,0)(75,30) \Line(75,0)(105,0) \GOval(75,60)(30,30)(0){0.411}
\end{picture}
+
\begin{picture}(40,45) (60,-3)
\SetScale{0.5}
\SetWidth{0.5} \SetColor{Black}
\Line(150,0)(195,0) \Line(195,0)(240,0) \Line(195,0)(165,35)
\Line(195,0)(225,35) \GOval(150,60)(30,30)(0){0.411}
\GOval(150,60)(30,30)(0){0.411} \GOval(240,60)(30,30)(0){0.411}
\end{picture}
\label{A0}
\ee

These contributions are given by

\be {\cal A}^0_{1\to n}={\cal A}_{1 \to n-1}+\sum_{n_1+n_2=n-1}{\cal
A}_{1\to n_1}{\cal A}_{1\to n_2}\ee

\item Secondly, BCFW is multiple-counting contributions of the form

  
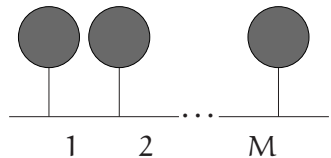
\begin{picture}(120,42) (0,0)
\SetScale{0.5}
    \SetWidth{0.5}
    \SetColor{Black}
    \Line(30,0)(158,0)
    \Line(188,0)(271,0)
    \Text(80,0)[lb]{\small{\Black{$\ldots$}}}
    \Line(60,0)(60,40)
    \Line(113,0)(113,40)
    \Line(233,0)(233,40)
    \GOval(60,60)(23,23)(0){0.411}
    \GOval(113,60)(23,23)(0){0.411}
    \GOval(233,60)(23,23)(0){0.411}
    \Text(36,-15)[lb]{\small{\Black{$1$}}}
    \Text(64,-15)[lb]{\small{\Black{$2$}}}
    \Text(105,-15)[lb]{\small{\Black{$M$}}}
  \end{picture}

or

  
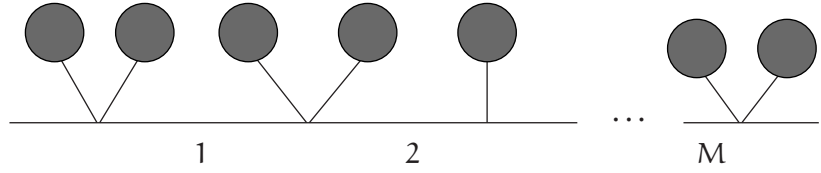
\begin{picture}(150,60) (15,-200)
    \SetWidth{0.5}
    \SetColor{Black}
    \Text(258,-184)[lb]{\small{\Black{$\ldots$}}}
    \SetWidth{0.5}
    \Line(284,-184)(335,-184)
    \Line(305,-184)(289,-162)
    \Line(306,-184)(323,-162)
    \GOval(323,-156)(11,11)(0){0.411}
    \GOval(289,-156)(11,11)(0){0.411}
    \Line(210,-184)(210,-150)
    \Line(63,-184)(47,-156)
    \Line(64,-184)(81,-156)
    \Line(142,-184)(120,-156)
    \Line(143,-184)(166,-156)
    \Line(30,-184)(244,-184)
    \GOval(47,-150)(11,11)(0){0.411}
    \GOval(81,-150)(11,11)(0){0.411}
    \GOval(120,-150)(11,11)(0){0.411}
    \GOval(165,-150)(11,11)(0){0.411}
    \GOval(210,-150)(11,11)(0){0.411}
    \Text(100,-200)[lb]{\small{\Black{$1$}}}
    \Text(290,-200)[lb]{\small{\Black{$M$}}}
    \Text(180,-200)[lb]{\small{\Black{$2$}}}
  \end{picture}

In fact we suggest that this over counting is exactly equal to the
multiplicity of propagator lines connecting particles $1$ and $n$.
\end{itemize}

To make our arguments more quantitative we start with the
Berends-Giele \cite{berends_giele} (or Dyson-Schwinger for ordered graphs 
\cite{Argyres:1992js},\cite{Draggiotis:1998gr})  recursive
equation for a generic theory with $3-$ and $4-$vertices.

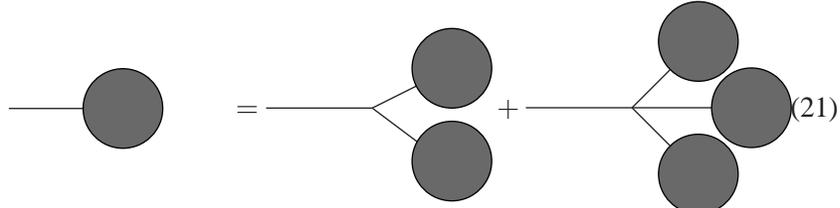
\be
\begin{picture}(85,40)(5,-8) 
\Line(7,-5)(50,-5)
    \GOval(50,-5)(15,15)(0){0.411}
\end{picture}
=
\begin{picture}(85,40)(10,32) 
\Line(10,35)(50,35)
    \Line(50,35)(70,45)
    \Line(50,35)(70,20)
    \GOval(80,50)(15,15)(0){0.411}
    \GOval(80,15)(15,15)(0){0.411}
\end{picture}
+
\begin{picture}(85,40)(15,47)
\Line(15,50)(55,50)
    \Line(55,50)(70,65)
    \Line(55,50)(85,50)
    \Line(55,50)(70,35)
    \GOval(80,75)(15,15)(0){0.411}
    \GOval(100,50)(15,15)(0){0.411}
    \GOval(80,25)(15,15)(0){0.411}
\end{picture}
\ee

\be {\cal A}_{1\to n}=\sum_{n_1+n_2=n} {\cal A}_{1\to n_1}{\cal
A}_{1\to n_2 }+  \sum_{n_1+n_2+n_3=n} {\cal A}_{1\to n_1}{\cal
A}_{1\to n_2 }{\cal A}_{1\to n_3}\ee

and we write the following equation

\be {\cal A}_{1\to n}={\cal A^{BCF}}_{1\to n}+{\cal A}^0_{1 \to
n}-D\ee

where ${\cal A}^0_{1 \to n}$ are the classes of diagrams in eq.\ref{A0} and the substracted 
term ${\cal D}$ accounts for the overcounting of diagrams. Since the overcounting relates to diagrams with
$2,3,\ldots$ propagators on the line between the first and the last leg, it can be seen that 
\be
{\cal D}=\sum_{M=3}^{n-1}(M-2) \sum_{k=0}^{M} \left( \begin{array}{c}
                       M\\
                       k
                     \end{array}\right)
 {\cal D}^n_{M+k} 
\label{counteq}
\ee
where $M-1$ is the number of propagators of the particular overcounted class of diagrams and 
\be {\cal D}^n_M=\sum_{n_1+\ldots +n_M=n} {\cal A}_{1\to n_1} \ldots
{\cal A}_{1\to n_M} \ee
counts the number of diagrams within that class. As an example for $n=5$ ($6$-leg gluon amplitude), the number
of Feynman graphs is ${\cal A}=38$, the number of BCFW graphs is ${\cal A^{BCF}}=29$, the number of graphs without 
any propagator between the first and the last leg is ${\cal A}^0=17$ so overcounting should give $D=8$. The following graphs
are overcounted
\be
\begin{picture}(200,50) (0,-15)

\SetScale{0.3}
    \SetWidth{0.5}
    \SetColor{Black}
    \Line(0,76.84)(102.46,76.84)
    \Line(25.61,76.84)(25.61,128.07)
    \Line(51.23,128.07)(51.23,76.84)
    \Line(76.84,128.07)(76.84,76.84)
    \Line(102.46,76.84)(128.07,76.84)
    \Line(102.46,76.84)(102.46,128.07)
    \Line(153.69,76.84)(281.76,76.84)
    \Line(179.3,76.84)(179.3,102.46)
    \Line(179.3,102.46)(153.69,128.07)
    \Line(179.3,102.46)(204.92,128.07)
    \Line(230.53,76.84)(230.53,128.07)
    \Line(256.15,76.84)(256.15,128.07)
    \Line(307.38,76.84)(435.45,76.84)
    \Line(332.99,76.84)(332.99,128.07)
    \Line(384.22,76.84)(384.22,102.46)
    \Line(384.22,102.46)(358.61,128.07)
    \Line(384.22,102.46)(409.84,128.07)
    \Line(435.45,76.84)(435.45,128.07)
    \Line(435.45,76.84)(461.07,76.84)
    \Line(486.68,76.84)(589.14,76.84)
    \Line(512.3,76.84)(512.3,128.07)
    \Line(537.91,128.07)(537.91,76.84)
    \Line(589.14,76.84)(589.14,102.46)
    \Line(589.14,102.46)(563.53,128.07)
    \Line(589.14,102.46)(614.76,128.07)
    \Line(589.14,76.84)(614.76,76.84)
    \Line(153.69,-25.61)(281.76,-25.61)
    \Line(179.3,-25.61)(153.69,25.61)
    \Line(179.3,-25.61)(204.92,25.61)
    \Line(230.53,25.61)(230.53,-25.61)
    \Line(256.15,25.61)(256.15,-25.61)
    \Line(307.38,-25.61)(461.07,-25.61)
    \Line(332.99,25.61)(332.99,-25.61)
    \Line(358.61,25.61)(384.22,-25.61)
    \Line(384.22,-25.61)(409.84,25.61)
    \Line(435.45,25.61)(435.45,-25.61)
    \Line(486.68,-25.61)(614.76,-25.61)
    \Line(512.3,-25.61)(512.3,25.61)
    \Line(537.91,25.61)(537.91,-25.61)
    \Line(563.53,25.61)(589.14,-25.61)
    \Line(589.14,-25.61)(614.76,25.61)
\Text(14,10)[lb]{\small{$(2)$}}  \end{picture}
\ee
with the first diagram doubly overcounted (it contains 3 propagators). That diagram
corresponds to the $M=4$ while the other graphs come from the $M=3$ term of eq.\ref{counteq}.

In the table below we give the results for up to $n=11$ particles.
\\

\begin{tabular}{|c|l|l|l|l|l|}
  \hline
  % after \\: \hline or \cline{col1-col2} \cline{col3-col4} ...
  $1\to n$ & ${\cal A}$ & ${\cal A^{BCF}}$ & ${\cal A}^0$
  & ${\cal A^{BCF}}+{\cal A}^0-{\cal A}$ & ${\cal D}$ \\
  \hline
  3   & 3      & 1      & 2     & 0      & 0 \\
  4   & 10     & 6      & 5     & 1      & 1 \\
  5   & 38     & 29     & 17    & 8      & 8 \\
  6   & 154    & 136    & 64    & 46     & 46 \\
  7   & 654    & 636    & 259   & 241    & 241 \\
  8   & 2871   & 2992   & 1098  & 1219   & 1219 \\
  9   & 12925  & 14190  & 4815  & 6080   & 6080 \\
  10  & 59345  & 67860  & 21659 & 30174  & 30174 \\
  11  & 276835 & 327080 & 99385 & 149630 & 149630 \\
  \hline
\end{tabular}
\newline
\newline
This analysis hints strongly towards the idea that a connection between Feynman diagrams and the BCFW decomposition might be achieved by grouping together BCFW (hatted) diagrams with the same chain structure along the main line (the same number of propagators hence the same multiplicity) but differently placed cuts:
The multiplicity of each group is equal to the number of propagators along the main line which in turns
equals the number of possible cuts (one for each propagator along the main line).

%--------------------------------------------------------------- Kinematical Id's
\section{Kinematical identities}

For any  function $\hat{f}(p;z)$ of $z$ that has only simple poles in $z$ and vanishes at $z\rightarrow \infty$, we can perform an expansion over its poles. In particular we have
\be
\hat{f}(p;z)=\sum_{j} \left[\hat{f}(p;z) (z-z_j)\right]_{z=z_j}\frac{1}{z-z_j}
\ee
where the sum is over all the simple poles $z_j$ of $\hat{f}(p;z)$.
This very general identity allows us to analytically continue the function $f(p)$ to the complex plane, make use of
the pole expansion and take the limit $z\rightarrow 0$ to return to the real axis, thus obtaining a relation between
$f(p)$ and the pole expansion of $\hat{f}(p;z)$.

Applying this to a momentum antenna we get
\be
\frac{1}{\hat{p}_{1}^2\hat{p}_{2}^2\ldots\hat{p}_{k}^2}=\sum_{j=1..k}\left[\frac{1}{\hat{p}_{1}^2\hat{p}_{2}^2\ldots
\hat{p}_{j-1}^2\hat{p}_{j+1}\ldots\hat{p}_{k}^2}\right]_{z=z_j}\frac{1}{\hat{p}_{j}^2}
\ee
with $z_j$ such that
\be
\hat{p}_{j}(z_j)^2=(p_j+z_j\epsilon)^2=p_j^2+2z_jp_j\mycdot\epsilon=0
\ee
where $p_1\ldots p_k$ are arbitrary off-shell four-momenta. In the next sections we will use the above identity
with $p_i$ being sums of on-shell momenta of the form $p_{1k}=p_1+p_2+\ldots+p_k$.

Taking the limit $z\rightarrow 0$ at both sides we have
\be
\frac{1}{p_{1}^2p_{2}^2\ldots p_{k}^2}=\sum_{j=1..k}\left[\frac{1}{\hat{p}_{1}^2\hat{p}_{2}^2\ldots
\hat{p}_{j-1}^2\hat{p}^2_{ j+1}\ldots\hat{p}_{k}^2}\right]_{z=z_j}\frac{1}{p_{j}^2}
\label{scalar_id}
\ee

Further more,
\be
\frac{z^{\rho}}{\hat{p}_{1}^2\hat{p}_{2}^2\ldots\hat{p}_{k}^2}=\sum_{j=1..k}\left[\frac{z^{\rho}}{\hat{p}_{1}^2\hat{p}_{2}^2\ldots
\hat{p}_{j-1}^2\hat{p}^2_{ j+1}\ldots\hat{p}_{k}^2}\right]_{z=z_j}\frac{1}{\hat{p}_{j}^2}
\label{kin}
\ee
which gives the very useful set of identities, valid for every $\rho<k$
\be
\sum_{j=1..k}\left[\frac{z^{\rho}}{\hat{p}_{1}^2\hat{p}_{2}^2\ldots
\hat{p}_{j-1}^2\hat{p}_{ j+1}\ldots\hat{p}_{k}^2}\right]_{z=z_j}\frac{1}{p_{j}^2}=0
\label{vector_id}
\ee

Finally, if $z=k$, the function in the left hand side of eq.\ref{kin} is no longer vanishing at $z\rightarrow \infty$.
Subtracting its limit at infinity, however, we have a new function that does, so
\be
\frac{z^{k}}{\hat{p}_{1}^2\hat{p}_{2}^2\ldots\hat{p}_{k}^2}-\lim_{z\rightarrow\infty}\frac{z^{k}}{\hat{p}_{1}^2\hat{p}_{2}^2\ldots\hat{p}_{k}^2}=\sum_{j=1..k}\left[\frac{z^{k}}{\hat{p}_{1}^2\hat{p}_{2}^2\ldots
\hat{p}_{j-1}^2\hat{p}_{ j+1}\ldots\hat{p}_{k}^2}\right]_{z=z_j}\frac{1}{\hat{p}_{j}^2}
\ee
and taking the limit $z\rightarrow 0$ we get

\be
-\lim_{z\rightarrow\infty}\frac{z^{k}}{\hat{p}_{1}^2\hat{p}_{2}^2\ldots\hat{p}_{k}^2}=\sum_{j=1..k}\left[\frac{z^{k}}{\hat{p}_{1}^2\hat{p}_{2}^2\ldots
\hat{p}_{j-1}^2\hat{p}_{ j+1}\ldots\hat{p}_{k}^2}\right]_{z=z_j}\frac{1}{p_{j}^2}
\ee
or
\be
\frac{-1}{\prod_{j=1}^{k}{2\epsilon\mycdot p_{j}}}=\sum_{j=1..k}\left[\frac{z^{k}}{\hat{p}_{1}^2\hat{p}_{2}^2\ldots
\hat{p}_{j-1}^2\hat{p}_{j+1}\ldots\hat{p}_{k}^2}\right]_{z=z_j}\frac{1}{p_{j}^2}
\label{full_id}
\ee
%--------------------------------------------------------------- gauge choice
\section{Choosing a gauge}
We have seen in section \ref{counting} that classes of Feynman
diagrams should somehow correspond to particular BCFW decompositions.
Using a particular gauge for the external gluons one can eliminate whole classes of Feynman diagrams.

A consistent definition of the external gluons' polarization vectors was given in eq.\ref{polarization_vectors_def}
where $\overline{p}$ is an auxiliary, null four-vector that can be chosen at will for every gluon as long as it is not 
parallel to $p$ itself.

We choose to use $\overline{p}_1=p_n$ and $\overline{p}_n=p_1$

\bea
\epsilon_{+1}^{\mu}=\frac{\ubar_{+n}\gamma^{\mu}u_{+1}}{\sqrt{2}\ubar_{+n}u_{-1}}=
\frac{\tl_{\dot{a}}(p_1)\overline{\sigma}^{\mu,\dot{a}a}\lambda_a(p_n)}{\sqrt{2}\bra n1 \ket}\\
\epsilon_{-n}^{\mu}=\frac{\ubar_{-1}\gamma^{\mu}u_{-n}}{\sqrt{2}\ubar_{-n}u_{+1}}=
-\frac{\tl_{\dot{a}}(p_1)\overline{\sigma}^{\mu,\dot{a}a}\lambda_a(p_n)}{\sqrt{2}\left[ n1 \right] }
\eea

A number of simplifications follow immediately. The product of $\epsilon_1$ with $\epsilon_n$ or $\epsilon$ (defined in 
eq.\ref{e_definition}) vanishes
\be
\fbox{$\epsilon_1\mycdot\epsilon_n=0=\epsilon\mycdot\epsilon_1=\epsilon\mycdot\epsilon_n$}
\label{e_id}
\ee

Moreover
\be
\epsilon_{+1}\mycdot p_n=0=\epsilon_{-n}\mycdot p_1.
\ee
As a consequence, any diagram in which the first and the last leg meet in a three-vertex vanishes.
\be
\begin{picture}(40,35) (0,3)
    \SetScale{0.4}
    \SetWidth{0.5}
    \SetColor{Black}
    \Line(0,-15)(45,15)
    \Line(45,15)(90,-15)
    \Line(45,15)(45,30)
    \GOval(45,45)(15,15)(0){0.411}
    \Text(-3,0)[lb]{\small{1}}
    \Text(35,0)[lb]{\small{n}}
  \end{picture}=0
\label{zerovertex1}
\ee
\vspace*{2pt}
where the blob denotes any Feynman diagram with the particular off-shell leg.

These simplifications are in the correct direction in view of the fact that there are no BCFW graphs
with the first and the last leg in the same vertex.  We still have to deal with the case when $1$ and $n$ are
attached in a four-vertex whose two other lines are off-shell (another situation that doesn't occur in a BCFW decomposition). 
We will see in the next section how to accommodate these diagrams.

Finally, we should note that the polarization vectors $\epsilon_{+1}^{\mu}$ and  $\epsilon_{-n}^{\mu}$
actually differ by a complex phase.They also differ by a complex factor from the 
shifting vector $\epsilon$ (see \ref{e_definition}).
Note that in any expression where all these three vectors appear in scalar products, one is allowed to freely interchange them,
without altering the result. In a very real sense there is actually only one polarization vector in this gauge.

Finally, the polarization vectors are now invariant under the shifting operation:
\be
p_1\rightarrow p_1+z\epsilon\;\;\;\; p_n\rightarrow p_n-z\epsilon
\ee
If we write
\be
p_1^{\mu}=\frac{1}{2}\ubar_{-1}\gamma^{\mu}u_{-1}\rightarrow\frac{1}{2}\ubar_{-1}\gamma^{\mu}u_{-1}+\frac{1}{2}z\ubar_{-1}\gamma^{\mu}u_{-n}
\ee
and 
\be
p_n^{\mu}=\frac{1}{2}\ubar_{-n}\gamma^{\mu}u_{-n}\rightarrow\frac{1}{2}\ubar_{-n}\gamma^{\mu}u_{-n}+\frac{1}{2}z\ubar_{-1}\gamma^{\mu}u_{-n}
\ee
we see that the shifting operator effectively sends
\bea
u_{-1}\rightarrow u_{-1}+z u_{-n}\nonumber\\
\ubar_{-n}\rightarrow \ubar_{-n}-z\ubar_{-1}
\eea
or, in terms of Weyl - Van der Waerden spinors 
\bea
\lambda_a(p_1)\rightarrow\lambda_{a}(p_1)+z\lambda_a(p_n)\nonumber\\
\tilde{\lambda}_{\dot{a}}(p_n)\rightarrow\tilde{\lambda}_{\dot{a}}(p_n)-z\tilde{\lambda}_{\dot{a}}(p_1)
\eea
As a result, the denominators of $\epsilon_{+1}$ and $\epsilon_{-n}$ become
\be
\bra n 1 \ket\rightarrow \bra n 1\ket + z \bra n n \ket=\bra n 1 \ket
\ee
and
\be
[n1]\rightarrow [n1]-z [11]=[n1]
\ee

The only restriction that we impose on the polarization vectors of the other gluons is that they remain 
invariant under the shifting operator. In case their assisting vectors $\overline{p}_i$ involve $p_1$ or $p_n$
this should be arranged in a way that preserves this invariance. 

This gauge choice explicitly eliminates any $z$-dependence from the polarization vectors. 
The shifting operation affects amplitudes only through the shift at the momenta four-vectors $p_1$ and $p_n$,
and the induced shift to the momenta of intermediate propagators. Polarization vectors and vertices that
do not carry $p_1$ and $p_n$ are left unchanged. Therefore, a diagram after shifting will be a 
complex function with poles coming from three-vertices or propagators. 

This is in contrast with a general gauge, where the
polarization vectors $\epsilon_{+1}$ and $\epsilon_{-n}$ get an extra factor involving $z$ in their
denominator. A pole expansion, in that case, would need to take into account (gauge dependent) poles 
coming from the gluon polarization vectors. 

Counting powers of $z$ in the general gauge shows (see the concluding discussion in \cite{proof}) 
that any diagram vanishes at the limit
$z\rightarrow\infty$. This is {\emph not} in general true in the gauge we are working. The $z\rightarrow\infty$ limit
of a diagram is used as a guide to group diagrams in classes where these limits cancel.

%--------------------------------------------------------------- diagrammatic proof
\section{Review of the BCFW decomposition}
According to the simplest version of the BCFW decomposition, the color amplitude $A(p_i;h_i)$,
where $h_i$ is the helicity of the $i$'th gluon, is equal to
\be
A(p_1^{h_1},\ldots,p_n^{h_n})=
\sum_{j=2}^{n-1}{ \sum_h{A(\hat{p}_1^{h_1},\ldots,p_j^{h_j},-\hat{P}_{1\ldots j}^h)
\frac{1}{P_{1\ldots j}^2}A(\hat{P}_{1\ldots j}^{-h},p_{j+1}^{h+1},\ldots, \hat{p}_n^{h_n})}}
\ee
The sum over $j$ extends over all partitions of the $n-2$ gluons in two groups. The amplitude
is recovered from the sum of $n-2$ decompositions in lower level amplitudes evaluated at
a particular $z=z_j$ such that $\hat{P}_{1\ldots j}(z_j)^2=0$.

 Let us call `hatted' diagrams graphs of the form
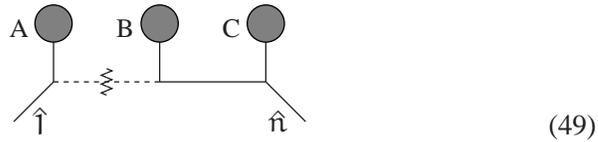
\be
\begin{picture}(120,60)(0,0)
\Line(5,5)(20,20)\DashLine(20,20)(60,20){2}\Line(60,20)(100,20)\Line(20,20)(20,35)\GCirc(20,42){7}{0.5}\PText(7,42)(0)[c]{A}
\Line(115,5)(100,20)\Line(60,20)(60,35)\GCirc(60,42){7}{0.5}\PText(47,42)(0)[c]{B}
\Line(100,20)(100,35)\GCirc(100,42){7}{0.5}\PText(87,42)(0)[c]{C}
\ZigZag(40,15)(40,25){2}{3}
\Text(15,5)[c]{$\hat{1}$}\Text(105,5)[c]{$\hat{n}$}
\end{picture}
\ee
denoting the corresponding Feynman graph where we have multiplied by $p_{1A}^2$,
we have performed the momenta shift,  evaluated it at some $z=z_i$ and divided back by $p_{1A}^2$. The value $z_i$ is defined by the demand that the `cut' propagator, carrying momentum $\hat{p}_{1A}$, vanishes: $\hat{p}_{1A}^2=0$.

In terms of `hatted' graphs, the BCFW decomposition consists in two blobs, one containing the first and one the last leg, the amplitude is hatted
and a cut is taken on the propagator that connects the two blobs. A sum over all partitions of $n$ in two integers is employed. Evidently, every particular BCFW decomposition over some propagator
$P_i$ is equivalent to the sum of all hatted diagrams with that particular propagator cut.

Thus the sum of all possible decompositions is
equivalent to the sum of all hatted diagrams with all possible cuts along the line of propagators that connects the first and the last leg of
the original color amplitude.

In what follows we will prove that the sum of all possible hatted diagrams is equivalent to
the sum of Feynman diagrams involved in the computation of the particular amplitude.

%--------------------------------------------------------------- diagrammatic proof
\section{The correspondence of hatted graphs to Feynman diagrams}
The special treatment of the two selected external legs $1$ and $n$ implies a special
classification of the various Feynman diagrams involved in the given amplitude. There are
classes of diagrams with zero, one,two,\ldots propagators on the line between the first and the last leg.
Every class consists of subclasses defined by the particular partition of the other $n-2$ external legs in
which they are connected to the main line.

A particular diagram with $n=20$ would be, for example, 
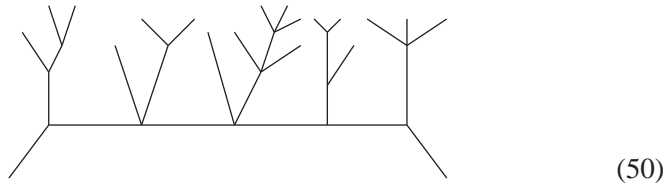
\be
\begin{picture}(165,65) (0,-5)
   \SetScale{1}
    \SetWidth{0.5}
    \SetColor{Black}
    \Line(0,-5)(15,15)
    \Line(15,15)(150,15)
    \Line(150,15)(165,-5)
    \Line(15,15)(15,35)
    \Line(15,35)(5,50)
    \Line(15,35)(20,45)
    \Line(20,45)(15,60)
    \Line(20,45)(25,60)
    %\Line(30,15)(30,30)
    %\Line(30,30)(30,50)
    \Line(50,15)(40,45)
    \Line(50,15)(60,45)
    \Line(60,45)(50,55)
    \Line(60,45)(70,55)
    \Line(85,15)(75,50)
    \Line(85,15)(95,35)
    \Line(95,35)(85,50)
    \Line(95,35)(100,50)
    \Line(95,35)(110,45)
    \Line(100,50)(95,60)
    \Line(100,50)(105,60)
    \Line(100,50)(110,55)
    \Line(150,15)(150,45)
    \Line(150,45)(165,55)
    \Line(150,45)(150,55)
    \Line(150,45)(135,55)
    \Line(120,15)(120,50)
    \Line(120,30)(130,45)
    \Line(120,50)(115,55)
    \Line(120,50)(125,55)
  \end{picture}
  \label{example}
\ee
and would correspond to the class
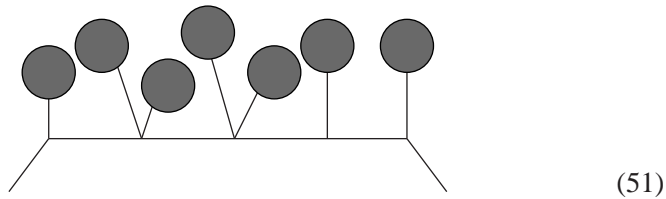
\be
\begin{picture}(165,70) (0,0)
    \SetScale{1}
    \SetWidth{0.5}
    \SetColor{Black}
    \Line(0,0)(15,20)
    \Line(15,20)(150,20)
    \Line(150,20)(165,0)
    \Line(15,20)(15,40)
    \Line(50,20)(40,50)
    \Line(50,20)(60,50)
    \Line(85,20)(75,55)
    \Line(85,20)(95,40)
    \Line(150,20)(150,50)
    \Line(120,20)(120,55)
    \GOval(150,55)(10,10)(0){0.411}
    \GOval(120,55)(10,10)(0){0.411}
    \GOval(100,45)(10,10)(0){0.411}
    \GOval(75,60)(10,10)(0){0.411}
    \GOval(60,40)(10,10)(0){0.411}
    \GOval(15,45)(10,10)(0){0.411}
    \GOval(35,55)(10,10)(0){0.411}
  \end{picture}
\label{subclass}
\ee
and more precicely to its subclass with $3,1,2,1,5,3,3$ on-shell legs in the corresponding
blobs\footnote{A blob with $m_i$ on-shell legs and one off-shell denotes the 
corresponding sum of color-ordered diagrams with one off-shell leg and $m_i$ on-shell legs.}. Let $m_1,m_2,\ldots,m_{k+1}$ be a particular partition of $n-2$, such that $\sum m_i=n-2$.
The blobs denote the full set of all possible color-ordered diagrams with one off-shell and
$m_i$ on-shell legs.

 Let us call this subclass of Feynman diagrams by
$F(m_1,\ldots,\tilde{m}_{j},\tilde{m}_{j+1},\ldots,m_{k+1})$ where the tildes on top of two consecutive $m$'s mark the presence of a four-vertex
attaching the corresponding blobs to the line.

\be
F(m_1,\ldots,\tilde{m}_{j},\tilde{m}_{j+1},\ldots,m_{k+1})=
\begin{picture}(285,30) (-5,-10)
    \SetScale{0.4}
    \SetWidth{0.5}
    \SetColor{Black}
    \Line(15,-17)(75,-17)
    \Line(75,-17)(75,28)
    \Line(75,-17)(90,-17)
    \Line(135,-17)(165,-17)
    \Line(165,-17)(195,-17)
    \Line(165,-17)(150,28)
    \Line(165,-17)(180,28)
    \Line(240,-17)(300,-17)
    \Line(255,-17)(255,28)
    \GOval(146,41)(15,15)(0){0.411}
    \GOval(185,41)(15,15)(0){0.411}
    \GOval(75,41)(15,15)(0){0.411}
    \GOval(255,40)(15,15)(0){0.411}
    \Text(39,-7)[lb]{\small{\Black{$\ldots$}}}
    \Text(81,-7)[lb]{\small{\Black{$\ldots$}}}
    \Text(23,28)[lb]{\small{$m_1$}}
    \Text(46,28)[lb]{\small{$\tilde{m}_j$}}
    \Text(62,28)[lb]{\small{$\tilde{m}_{j+1}$}}
    \Text(95,28)[lb]{\small{$m_n$}}
    \Text(10,-18)[lb]{\small{$1$}}
    \Text(120,-18)[lb]{\small{$n$}}

  \end{picture}
\ee
The particular subclass of diagrams in eq.\ref{subclass} would then be called $F(3,\tilde{1},\tilde{2},\tilde{1},\tilde{5},3,3)$ and eq.\ref{example}
would be one of the actual Feynman diagrams contained in that subclass.

In the gauge we are working, the hatting operator would leave the contribution of the blobs, in such a diagram, invariant since the
momentum flowing into the blob doesn't contain $p_1$ or $p_n$. It would only
affect the propagators and vertices along the main line. After the shifting of the momenta a cut
can be placed on any of the propagators on the line, thus fixing the value of $z$ at which the residue should be evaluated. We shall denote
such a graph with a cut on the $j$'th propagator by
\be
H(m_1,\ldots,m_{k+1};j)=\left[\hat{F}(m_1,\ldots,m_{k+1})\hat{p}_{1j}^2\right]_{z=z_i} \frac{1}{p_{1j}^2}
\ee

Here and in what follows $p_{1A_i}$ denotes the sum of external momenta
\be
p_{1A_i}=p_1+p_{A_i}+p_{A_2}+\ldots+p_{A_i}
\ee
with
\be
p_{A_i}=\sum_{k=1}^{m_i}{p_k}
\ee

A closer look at the "hatted" diagram will reveal that  hatting affects, through the momenta $p_1$ and $p_n$, only the
three-vertices along the main line. The four-vertices, not carrying any momentum, remain unchanged.
The effect on the three-vertices is included in the "hatted" YM three-gluon vertices,
which we denote with a crossed white blob in the figures. Each of these can be decomposed in two pieces:
\bea
\begin{picture}(30,30) (20,-5)
    \SetScale{0.3}
    \SetWidth{0.5}
    \SetColor{Black}
    \Line(15,-4)(90,-4)
    \Line(90,-4)(90,56)
    \Line(90,-4)(150,-4)
    \Line(150,-4)(165,-4)
    \GOval(90,56)(15,15)(0){0.411}
    \COval(90,-4)(2.83,2.83)(45.0){Black}{White}
\Line(88.59,-5.41)(91.41,-2.59)\Line(88.59,-2.59)(91.41,-5.41)
    \COval(90,-4)(2.83,2.83)(45.0){Black}{White}
\Line(88.59,-5.41)(91.41,-2.59)\Line(88.59,-2.59)(91.41,-5.41)
    \COval(89,-3)(11.31,11.31)(-45.0){Black}{White}
\Line(94.66,-8.66)(83.34,2.66)\Line(83.34,-8.66)(94.66,2.66)
  \end{picture}
&=&\hat{V}_{\mu\nu\rho}=g_{\mu\nu}(\hat{p}_1-p_2)_{\rho}+g_{\nu\rho}(p_2-\hat{p}_3)_{\mu}+g_{\rho\mu}(\hat{p}_3-\hat{p}_1)_{\nu}\nonumber \\
&=&g_{\mu\nu}(p_1-p_2)_{\rho}+g_{\nu\rho}(p_2-p_3)_{\mu}+g_{\rho\mu}(p_3-p_1)_{\nu}\nonumber\\
& &-z\epsilon^{\sigma}(2 g_{\mu\rho} g_{\nu\sigma}-
g_{\mu\sigma} g_{\nu\rho}-g_{\mu\nu} g_{\sigma\rho})\nonumber\\
&=&V_{\mu\nu\rho}-z\epsilon^{\sigma}(2 g_{\mu\rho} g_{\nu\sigma}-
g_{\mu\sigma} g_{\nu\rho}-g_{\mu\nu} g_{\sigma\rho})\nonumber\\
&=&
\begin{picture}(32,30) (10,-5)
    \SetScale{0.3}
    \SetWidth{0.5}
    \SetColor{Black}
    \Line(15,-4)(90,-4)
    \Line(90,-4)(90,56)
    \Line(90,-4)(150,-4)
    \Line(150,-4)(165,-4)
    \GOval(90,56)(15,15)(0){0.411}
  \end{picture}
\;\;\; +\;\;\;
 \begin{picture}(30,30) (10,-5)
    \SetScale{0.3}
    \SetWidth{0.5}
    \SetColor{Black}
    \Line(15,-4)(90,-4)
    \Line(90,-4)(90,56)
    \Line(90,-4)(150,-4)
    \Line(150,-4)(165,-4)
    \GOval(90,56)(15,15)(0){0.411}
    \Line(90,-4)(90,-45)
    \Text(30,-15)[lb]{\small{$\epsilon_{\sigma}$}}
    \COval(90,-4)(8.83,8.83)(60.0){Black}{White}
  \end{picture}
  \label{vertices}
\eea
where
\be
\begin{picture}(45,30) (10,-5)
    \SetScale{0.3}
    \SetWidth{0.5}
    \SetColor{Black}
    \Line(15,-4)(90,-4)
    \Line(90,-4)(90,56)
    \Line(90,-4)(150,-4)
    \Line(150,-4)(165,-4)
    \GOval(90,56)(15,15)(0){0.411}
    \Line(90,-4)(90,-45)
    \Text(30,-15)[lb]{\small{$\epsilon_{\sigma}$}}
    \COval(90,-4)(8.83,8.83)(60.0){Black}{White}
  \end{picture}=-z V_{\mu\nu\rho\sigma}
\ee
with $V_{\mu\nu\rho\sigma}$ the QCD four-vertex.

Let us concentrate for the moment on the class of diagrams with the smallest number of propagators 
along the line between the first and the last leg: diagrams with no propagators along the line 
can have a three or a four-vertex joining these two legs. The whole class of diagrams with
a three-vertex joining the first and the last leg vanishes identically in the gauge we are using. The remaning class
with one four-vertex will be dealt with soon.

Next, we have the class with one propagator along the main line.

If we write $J_Q^{\mu}$ for the current coming from the blob $Q$, we have
\bea
H(m_1,m_2;1)&\equiv&
\begin{picture}(80,40)(0,17)
\Line(5,5)(20,20)\Line(20,20)(10,30)\GCirc(7,33){7}{0.5}\PText(20,33)(0)[c]{A}
\Line(75,5)(60,20)\Line(60,20)(70,30)\GCirc(73,33){7}{0.5}\PText(60,33)(0)[c]{B}
\DashLine(20,20)(60,20){2}
\ZigZag(40,15)(40,25){2}{2}
\end{picture}\nonumber\\[3mm] 
&=&\left[\epsilon_{1\mu}J_{A\nu}\hat{V}^{\mu\nu\rho}\hat{V}_{\rho\kappa\lambda}J_B^{\kappa}\epsilon_n^{\lambda}\right]_{z_0}\frac{1}{p_{1A}^2}\nonumber\\
&=&\left[\epsilon_{1\mu}J_{A\nu}(V^{\mu\nu\rho}-z\epsilon_{\sigma}V^{\sigma\mu\nu\rho})
(V_{\rho\kappa\lambda}-z\epsilon^{\tau}V_{\tau\rho\kappa\lambda})J_B^{\kappa}\epsilon_n^{\lambda}\right]_{z_0}\frac{1}{p_{1A}^2}\nonumber\\
&=&\left[\epsilon_{1\mu}J_{A\nu}V^{\mu\nu\rho}V_{\rho\kappa\lambda}J_B^{\kappa}\epsilon_n^{\lambda}
-z\epsilon_{1\mu}J_{1\nu}\epsilon_{\sigma}V^{\sigma\mu\nu\rho}V_{\rho\kappa\lambda}J_B^{\kappa}\epsilon_n^{\lambda}\nonumber\right.\\
& &\left. -z\epsilon_{1\mu}J_{A\nu}V^{\mu\nu\rho}\epsilon^{\tau}V_{\tau\rho\kappa\lambda}J_B^{\kappa}\epsilon_n^{\lambda}\right.\nonumber\\
& &
\left.+z^2\epsilon_{1\mu}J_{A\nu}\epsilon_{\sigma}V^{\sigma\mu\nu\rho}\epsilon^{\tau}V_{\tau\rho\kappa\lambda}J_B^{\kappa}\epsilon_n^{\lambda}
\right]_{z_0}\frac{1}{p_{1A}^2}\nonumber\\
\eea
The first, $z$-independent term in the bracket equals the Feynman diagram with one propagator.
The last term in the bracket vanishes due to eq.\ref{e_id} and we get
\bea
H(m_1,m_2;1)&\equiv&
\begin{picture}(80,40)(0,17)
\Line(5,5)(20,20)\Line(20,20)(10,30)\GCirc(7,33){7}{0.5}\PText(20,33)(0)[c]{A}
\Line(75,5)(60,20)\Line(60,20)(70,30)\GCirc(73,33){7}{0.5}\PText(60,33)(0)[c]{B}
\DashLine(20,20)(60,20){2}
\ZigZag(40,15)(40,25){2}{2}
\end{picture}=\nonumber\\[3mm] 
&=&\begin{picture}(80,40)(0,17)
\Line(5,5)(20,20)\Line(20,20)(10,30)\GCirc(7,33){7}{0.5}\PText(20,33)(0)[c]{A}
\Line(75,5)(60,20)\Line(60,20)(70,30)\GCirc(73,33){7}{0.5}\PText(60,33)(0)[c]{B}
\Line(20,20)(60,20)
\end{picture}
-\left[2(\epsilon\mycdot p_{1A})(\epsilon_1\mycdot J_A)(\epsilon_n\mycdot J_B)\right]_{z_0}\frac{1}{p_{1A}^2}\nonumber
\eea

with 
\be
z_0=-\frac{p_{1A}^2}{2(p_{1A}\epsilon)}
\ee
where we have used momentum conservation and our freedom to interchange the polarization vectors.
So
\bea
H(m_1,m_2;1)&\equiv&
\begin{picture}(80,40)(0,17)
\Line(5,5)(20,20)\Line(20,20)(10,30)\GCirc(7,33){7}{0.5}\PText(20,33)(0)[c]{A}
\Line(75,5)(60,20)\Line(60,20)(70,30)\GCirc(73,33){7}{0.5}\PText(60,33)(0)[c]{B}
\DashLine(20,20)(60,20){2}
\ZigZag(40,15)(40,25){2}{2}
\end{picture}\nonumber\\[3mm] 
&=&\begin{picture}(80,40)(0,17)
\Line(5,5)(20,20)\Line(20,20)(10,30)\GCirc(7,33){7}{0.5}\PText(20,33)(0)[c]{A}
\Line(75,5)(60,20)\Line(60,20)(70,30)\GCirc(73,33){7}{0.5}\PText(60,33)(0)[c]{B}
\Line(20,20)(60,20)
\end{picture}
+
\begin{picture}(40,40)(0,17)
\Line(5,5)(20,20)\Line(20,20)(10,30)\GCirc(7,33){7}{0.5}
\Line(35,5)(20,20)\Line(20,20)(30,30)\GCirc(33,33){7}{0.5}
\end{picture}
\label{one_prop}
\eea
or 
\be
H(m_1,m_2;1)=F(m_1,m_2)+F(\tilde{m}_1,\tilde{m}_2)
\ee

In the case that the blobs $A$ and $B$ of the first diagram in eq.\ref{one_prop} contain exactly one on-shell leg, the
BCFW decomposition leads to two hatted QCD 3-vertices. 
We have verified that the usual
QCD 3-vertex evaluated at the hatted kinematics reproduces
the formulas 
\be
A(\hat{1}^+,2^-,\hat{q}^+)=\sqrt{2}\frac{[\hat{q}\hat{1}]^3}{[\hat{1}2][2\hat{q}]}
\ee
and 
\be
A(\hat{1}^+,2^+,\hat{q}^-)=\sqrt{2}\frac{[\hat{1}\hat{2}]^3}{[2\hat{q}][\hat{q}1]}
\ee
where $p_q=-p_1-p_2$ is the off-shell leg and $\hat{p}_q=-p_1-p_2-z\epsilon$ is its on-shell continuation 
\footnote{The other polarizations ($A(\hat{1}^+,2^+,\hat{q}^+),A(\hat{1}^+,2^-,\hat{q}^-)$) 
can be shown to vanish.}. The corresponding formulas involving the last leg (which has negative helicity) work 
similarly but include angle bracket spinor products. 

Proceeding to graphs that contain two propagators between the first and the last leg or graphs that contain
one propagator with one four-vertex attached to it, we have
\bea
\begin{picture}(120,60)(0,17)
\Line(5,5)(20,20)\DashLine(20,20)(60,20){2}\Line(60,20)(100,20)\Line(20,20)(20,35)\GCirc(20,42){7}{0.5}\PText(7,42)(0)[c]{A}
\Line(115,5)(100,20)\Line(60,20)(60,35)\GCirc(60,42){7}{0.5}\PText(47,42)(0)[c]{B}
\Line(100,20)(100,35)\GCirc(100,42){7}{0.5}\PText(87,42)(0)[c]{C}
\ZigZag(40,15)(40,25){2}{2}
%\PText(15,8)(0)[c]{1}\PText(105,8)(0)[c]{n}
\end{picture}
&+&
\begin{picture}(120,60)(0,17)
\Line(5,5)(20,20)\Line(20,20)(60,20)\DashLine(60,20)(100,20){2}\Line(20,20)(20,35)\GCirc(20,42){7}{0.5}\PText(7,42)(0)[c]{A}
\Line(115,5)(100,20)\Line(60,20)(60,35)\GCirc(60,42){7}{0.5}\PText(47,42)(0)[c]{B}
\Line(100,20)(100,35)\GCirc(100,42){7}{0.5}\PText(87,42)(0)[c]{C}
\ZigZag(80,15)(80,25){2}{2}
%\PText(15,8)(0)[c]{1}\PText(105,8)(0)[c]{n}
\end{picture}
\\
+
\begin{picture}(80,60)(0,17)
\Line(5,5)(20,20)\Line(20,20)(10,40)\GCirc(7,43){7}{0.5}\PText(20,43)(0)[c]{A}
\Line(75,5)(60,20)\Line(60,20)(70,40)\GCirc(73,43){7}{0.5}\PText(60,43)(0)[c]{C}
\Line(60,20)(50,40)\GCirc(47,43){7}{0.5}\PText(35,43)(0)[c]{B}
\DashLine(20,20)(60,20){2}\ZigZag(40,15)(40,25){2}{2}
\end{picture}
&+&
\begin{picture}(80,60)(0,17)
\Line(5,5)(20,20)\Line(20,20)(10,40)\GCirc(7,43){7}{0.5}\PText(20,43)(0)[c]{A}
\Line(75,5)(60,20)\Line(60,20)(70,40)\GCirc(73,43){7}{0.5}\PText(60,43)(0)[c]{C}
\Line(20,20)(35,40)\GCirc(32,43){7}{0.5}\PText(45,43)(0)[c]{B}
\DashLine(20,20)(60,20){2}\ZigZag(40,15)(40,25){2}{2}
\end{picture}
\nonumber\\
=
\begin{picture}(120,60)(0,17)
\Line(5,5)(20,20)\Line(20,20)(100,20)\Line(20,20)(20,35)\GCirc(20,42){7}{0.5}\PText(7,42)(0)[c]{A}
\Line(115,5)(100,20)\Line(60,20)(60,35)\GCirc(60,42){7}{0.5}\PText(47,42)(0)[c]{B}
\Line(100,20)(100,35)\GCirc(100,42){7}{0.5}\PText(87,42)(0)[c]{C}
\end{picture}
&+&
\begin{picture}(80,60)(0,17)
\Line(5,5)(20,20)\Line(20,20)(10,40)\GCirc(7,43){7}{0.5}\PText(20,43)(0)[c]{A}
\Line(75,5)(60,20)\Line(60,20)(70,40)\GCirc(73,43){7}{0.5}\PText(60,43)(0)[c]{C}
\Line(60,20)(50,40)\GCirc(47,43){7}{0.5}\PText(35,43)(0)[c]{B}
\Line(20,20)(60,20)
\end{picture}
+
\begin{picture}(80,60)(0,17)
\Line(5,5)(20,20)\Line(20,20)(10,40)\GCirc(7,43){7}{0.5}\PText(20,43)(0)[c]{A}
\Line(75,5)(60,20)\Line(60,20)(70,40)\GCirc(73,43){7}{0.5}\PText(60,43)(0)[c]{C}
\Line(20,20)(35,40)\GCirc(32,43){7}{0.5}\PText(45,43)(0)[c]{B}
\Line(20,20)(60,20)
\end{picture}
\label{two_propagators}
\eea
The expressions that correspond to the first two hatted graphs on the left hand side
are quadradic in $z$ and are evaluated in different values of $z$, $z_1$ and $z_2$.
\be
\begin{picture}(120,60)(0,17)
\Line(5,5)(20,20)\DashLine(20,20)(60,20){2}\Line(60,20)(100,20)\Line(20,20)(20,35)\GCirc(20,42){7}{0.5}\PText(7,42)(0)[c]{A}
\Line(115,5)(100,20)\Line(60,20)(60,35)\GCirc(60,42){7}{0.5}\PText(47,42)(0)[c]{B}
\Line(100,20)(100,35)\GCirc(100,42){7}{0.5}\PText(87,42)(0)[c]{C}
\ZigZag(40,15)(40,25){2}{2}
%\PText(15,8)(0)[c]{1}\PText(105,8)(0)[c]{n}
\end{picture}=\left[H_0\frac{1}{\hat{p}_{1B}^2}+H_1 \frac{z}{\hat{p}_{1B}^2}+ H_2 \frac{z^2}{\hat{p}_{1B}^2}\right]_{z=z_1}\frac{1}{p_{1A}^2}
\ee
with $p_{1B}=p_1+p_A+p_B$
\be
\begin{picture}(120,60)(0,17)
\Line(5,5)(20,20)\Line(20,20)(60,20)\DashLine(60,20)(100,20){2}\Line(20,20)(20,35)\GCirc(20,42){7}{0.5}\PText(7,42)(0)[c]{A}
\Line(115,5)(100,20)\Line(60,20)(60,35)\GCirc(60,42){7}{0.5}\PText(47,42)(0)[c]{B}
\Line(100,20)(100,35)\GCirc(100,42){7}{0.5}\PText(87,42)(0)[c]{C}
\ZigZag(80,15)(80,25){2}{2}
%\PText(15,8)(0)[c]{1}\PText(105,8)(0)[c]{n}
\end{picture}=\left[H_0\frac{1}{\hat{p}_{1A}^2}+H_1 \frac{z}{\hat{p}_{1A}^2}+ H_2 \frac{z^2}{\hat{p}_{1A}^2}\right]_{z=z_2}\frac{1}{p_{1B}^2}
\ee

The $z$-independent term of all four graphs gives the Feynman graphs of the right hand side of eq.\ref{two_propagators},
due to the kinematic identity eq.\ref{scalar_id}.

Moreover, the part linear  in $z$ of the first two graphs
cancels identically due to eq.\ref{vector_id} with $\rho=1$. The quadratic part of the first two graphs
cancels exactly the part linear  in $z$  of the two last graphs on the left hand side of eq.\ref{two_propagators}.

The generalization of the above mechanism goes as follows: 
Each hatted diagram with $v_3$ three-vertices along the main line can be decomposed in a sum of sub-diagrams having
$0$,$1$, $\ldots$, $v_3$ white-blob vertices corresponding to $0$, $1$, $\ldots$, $v_3$ powers of $z$ (evaluated at some $z_i$). If the
diagram also has $v_4$ four-vertices, the number of propagators along the line is $v_3+v_4-1$.

The sum over all possible cuts for the particular diagram can then be written as
\bea
\lefteqn{\sum_{j=1}^{v_3+v_4-1}{H(m_1\ldots m_k;j)}=}\nonumber\\
&=&\sum_{j=1}^{v_3+v_4-1}{\left[(H_0+zH_1+z^2H_2+\ldots H_{v_3}z^{v_3})\prod_{q\neq j}\frac{1}{\hat{p}_{1q}^2}\right]_{z=z_j}\frac{1}{p_{1j}^2}}
\eea
or
\bea
\sum_{j=1}^{v_3+v_4-1}{H(m_1\ldots m_k;j)}&=&H_0\sum_{j=1}^{v_3+v_4-1}{\left[\frac{1}{\prod_{q\neq j}\hat{p}_{1q}^2}\right]_{z=z_j}\frac{1}{p_{1j}^2}} \nonumber\\
&&+H_1\sum_{j=1}^{v_3+v_4-1}{\left[\frac{z}{\prod_{q\neq j}\hat{p}_{1q}^2}\right]_{z=z_j}\frac{1}{p_{1j}^2}} \nonumber\\
&&+H_2\sum_{j=1}^{v_3+v_4-1}{\left[\frac{z^2}{\prod_{q\neq j}\hat{p}_{1q}^2}\right]_{z=z_j}\frac{1}{p_{1j}^2}}\nonumber\\
&&+\ldots \nonumber\\
& &+ H_{v_3}\sum_{j=1}^{v_3+v_4-1}{\left[\frac{z^{v_3}}{\prod_{q\neq j}\hat{p}_{1q}^2}\right]_{z=z_j}\frac{1}{p_{1j}^2}}
\label{eqH}
\eea
Due to the set of identities eq.\ref{vector_id}, all terms involving $z^{\lambda}$ with $0<\lambda<v_3+v_4-1$ vanish identically.

Moreover it is easy to see that the $z$-independent term involving $H_0$ will give the corresponding Feynman diagram
with the help of the identity eq.\ref{scalar_id}.

Let us further distinguish among three cases: diagrams with no four-vertex on the line ($v_3=k+1$), with one four-vertex ($v_3=k$)
and with more than one four-vertex ($v_3<k$). In the third case the right hand side of eq.\ref{eqH} reduces to the
$z$-independent term corresponding to the Feynman diagram. All terms involving $z$ vanish. In other words the sum over all possible cuts
of the hatted diagrams with two or more four-vertices on the line,
 is equal to the corresponding Feynman diagrams.

 In the first two cases the sum of all hatted diagrams over all possible
cuts gives the corresponding Feynman diagrams plus a number of terms. We will now show that these terms cancel exactly
each other due to the structure of the YM vertices.

Diagrams with no four vertex on the line have $v_3=k+1$ three-vertices. They are of the form
\be
H(m_1,\ldots,m_j,m_{j+1},\ldots,m_{k+1};j)=
\begin{picture}(285,30)(-5,-10)
    \SetScale{0.4}
    \SetWidth{0.5}
    \SetColor{Black}
    \Line(15,-17)(75,-17)
    \Line(75,-17)(75,28)
    \Line(75,-17)(90,-17)
    \Line(135,-17)(150,-17)
    \DashLine(150,-17)(180,-17){1}
    \ZigZag(165,-27)(165,-10){2}{3}
    \Line(150,-17)(150,28)
    \Line(180,-17)(180,28)
    \Line(180,-17)(195,-17)
    \Line(240,-17)(300,-17)
    \Line(255,-17)(255,28)
    \GOval(146,41)(15,15)(0){0.411}
    \GOval(185,41)(15,15)(0){0.411}
    \GOval(75,41)(15,15)(0){0.411}
    \GOval(255,40)(15,15)(0){0.411}
    \Text(39,-7)[lb]{\small{\Black{$\ldots$}}}
    \Text(81,-7)[lb]{\small{\Black{$\ldots$}}}
    \Text(23,28)[lb]{\small{$m_1$}}
    \Text(46,28)[lb]{\small{$m_j$}}
    \Text(62,28)[lb]{\small{$m_{j+1}$}}
    \Text(95,28)[lb]{\small{$m_n$}}
    \Text(10,-18)[lb]{\small{$1$}}
    \Text(120,-18)[lb]{\small{$n$}}

  \end{picture}
\ee
The highest order term involves $H_{k+1}$ that
vanishes identically: it corresponds to diagrams with only crossed white blobs  on the line, hence it consists of
contractions of $\epsilon$ with itself or $\epsilon_1,\epsilon_n$. The next to leading order term, $H_{k}$ is a sum of
terms with one QCD vertex and $k$ crossed white-blob vertices.

We have
\bea
\sum_{j=1}^{k}{H(m_1\ldots m_{k+1};j)}&=&F(m_1\ldots m_{k+1})+\nonumber\\
& &+H_{k}\sum_{j=1}^{k}{\left[\frac{z^{k}}{\prod_{q\neq j}\hat{p}_{1q}^2}\right]_{z=z_j}\frac{1}{p_{1j}^2}}
\eea
and using the kinematical identity eq.\ref{full_id} we get
\be
\sum_{j=1}^{k}{H(m_1\ldots m_{k+1};j)}=F(m_1\ldots m_{k+1})-
H_{k}\frac{1}{\prod_{j=1}^{k}{2\epsilon\mycdot p_{1j}}}
\ee

Let us denote $H_{k,r}$ the term where the QCD vertex is coming from the $r$'th
hatted vertex.

$H_{k,r}$ is the numerator of a diagram containing blobs connected with the main line by white-blob
vertices $\epsilon^c M_{\mu\nu\rho c}$ everywhere except the QCD $r$'th vertex. A generic piece on the line will contribute by
\bea
\begin{picture}(50,30) (30,-2)
    \SetScale{0.4}
    \SetWidth{0.5}
    \SetColor{Black}
    \Line(15,-1)(90,-1)
    \Line(90,-1)(90,59)
    \Line(90,-1)(150,-1)
    \Line(150,-1)(165,-1)
    \GOval(90,59)(15,15)(0){0.411}
\Line(90,-1)(90,-35)
\Text(38,-15)[lb]{\small{\Black{$\epsilon$}}}
    \COval(90,-1)(7,7)(0){Black}{White}
    \Text(-10,-1)[lb]{\small{\Black{$\ldots$}}}
    \Text(70,-1)[lb]{\small{\Black{$\ldots$}}}
    \Text(5,4)[lb]{\small{\Black{$\mu$}}}
    \Text(60,4)[lb]{\small{\Black{$\rho$}}}
    \Text(45,25)[lb]{\small{\Black{$A_i$}}}
  \end{picture}
  &\equiv& M(A_i)_{\mu\rho}\nonumber=-J_{A_i}^{\nu}\epsilon^c (2g_{c\nu}g_{\mu\rho}-g_{c\mu}g_{\nu\rho}-g_{c\rho}g_{\mu\nu})\nonumber\\
&=&-(2 (J_{A_i}\mycdot\epsilon)g_{\mu\rho}-J_{A_i,\mu}\epsilon_{\rho}-J_{A_i,\rho}\epsilon_{\mu})
\eea

where $J_{A_i}^{\nu}$ is the current (subamplitude) coming from the $i$'th blob.
The first such white-blob vertex, when contracted with $\epsilon_1^{\mu}$, gives
\be
\begin{picture}(50,30) (30,-2)
    \SetScale{0.4}
    \SetWidth{0.5}
    \SetColor{Black}
    \Line(15,-1)(90,-1)
    \Line(90,-1)(90,59)
    \Line(90,-1)(150,-1)
    \Line(150,-1)(165,-1)
    \GOval(90,59)(15,15)(0){0.411}
\Line(90,-1)(90,-35)
\Text(38,-15)[lb]{\small{\Black{$\epsilon$}}}
    \COval(90,-1)(7,7)(0){Black}{White}

    \Text(70,-1)[lb]{\small{\Black{$\ldots$}}}
    \Text(5,4)[lb]{\small{\Black{$1$}}}
    \Text(60,4)[lb]{\small{\Black{$\rho$}}}
    \Text(45,25)[lb]{\small{\Black{$A_1$}}}
  \end{picture}
  =\epsilon_1^{\mu}M(A_1)_{\mu\rho}=-(\epsilon_1\mycdot J_{A_1})\epsilon_{\rho}
\ee
It is easy to see that a chain of consecutive white-blob vertices gives
\be
\begin{picture}(130,30) (30,-2)
    \SetScale{0.4}
    \SetWidth{0.5}
    \SetColor{Black}
    \Line(15,-1)(90,-1)
    \Line(90,-1)(90,59)
    \Line(90,-1)(150,-1)
    \Line(150,-1)(165,-1)
    \GOval(90,59)(15,15)(0){0.411}
\Line(90,-1)(90,-35)
\Text(38,-15)[lb]{\small{\Black{$\epsilon$}}}
    \COval(90,-1)(7,7)(0){Black}{White}
    \Line(120,-1)(195,-1)
    \Line(195,-1)(195,59)
    \Line(195,-1)(225,-1)
    %\Line(255,-1)(270,-1)
    \GOval(195,59)(15,15)(0){0.411}
\Line(195,-1)(195,-35)
\Text(80,-15)[lb]{\small{\Black{$\epsilon$}}}
    \COval(195,-1)(7,7)(0){Black}{White}
    \Line(275,-1)(300,-1)
    \Line(300,-1)(300,59)
    \Line(300,-1)(360,-1)
    \Line(360,-1)(375,-1)
    \GOval(300,59)(15,15)(0){0.411}
\Line(300,-1)(300,-35)
\Text(125,-15)[lb]{\small{\Black{$\epsilon$}}}
    \COval(300,-1)(7,7)(0){Black}{White}
    \Text(95,-1)[lb]{\small{\Black{$\ldots$}}}
    \Text(45,25)[lb]{\small{\Black{$A_1$}}}
    \Text(88,25)[lb]{\small{\Black{$A_2$}}}
    \Text(129,25)[lb]{\small{\Black{$A_s$}}}
    \Text(5,4)[lb]{\small{\Black{$1$}}}
    \Text(150,4)[lb]{\small{\Black{$\rho$}}}
  \end{picture}
  =(-1)^{\rho}(\epsilon_1\mycdot J_{A_1})(\epsilon\mycdot J_{A_2})\ldots(\epsilon\mycdot J_{A_s})\epsilon_{\rho}
\ee
Since $H_{k,r}$ has $k$ such vertices we get
\bea
H_{k,r}&=&(-1)^k\prod_{j\neq r}(\epsilon\mycdot J_{A_j}) \epsilon_1^{\mu}J_{A_r}^{\nu}\epsilon_n^{\rho}\times\nonumber\\
& &\times(g_{\mu\nu}(p_{1,r-1}-p_r)_{\rho}
+g_{\nu\rho}(p_r+p_{1,r})_{\mu}
+g_{\rho\mu}(-p_{1,r}-p_{1,r-1})_{\nu})\nonumber\\
&=&(-1)^k\prod_{q\neq r}(\epsilon\mycdot J_{A_q}) (\epsilon_n\mycdot J_{A_r})((p_{1,r-1}+p_{1,r})\mycdot \epsilon_1)
\eea
where repeated use was made of our ability to interchange $\epsilon_1,\epsilon_n$ and $\epsilon$. Note that for the boundary terms,
$r=1$ and $r=k+1$ the above formula must be understood with the definitions $p_{1,0}=p_{1,k+2}=0$.

Summing over $r$  we regain $H_k$, and
\bea
\lefteqn{\sum_{j=1}^{k}{H(m_1\ldots m_{k+1};j)}=F(m_1\ldots m_{k+1})+}\nonumber\\
& &+(-1)^{k+1}\frac{1}{\prod_{j}{2\epsilon\mycdot p_{1j}}}\sum_{r}(\epsilon_1\mycdot J_{A_1})\ldots(\epsilon_n\mycdot J_{A_n})[(p_{1,r-1}+p_{1,r})\mycdot
\epsilon]\nonumber\\
&=&F(m_1\ldots m_k)+\nonumber\\
& &+(-1)^{k+1}
(\epsilon_1\mycdot J_{A_1})\ldots(\epsilon_n\mycdot J_{A_n})\sum_{r}
{\frac{2p_{1,r}\mycdot\epsilon}{(2\epsilon\mycdot p_{11})(2\epsilon\mycdot p_{12})\ldots (2\epsilon\mycdot p_{1k})}}\nonumber\\
&=&F(m_1\ldots m_k)+\nonumber\\
& &(-1)^{k+1}
(\epsilon_1\mycdot J_{A_1})\ldots(\epsilon_n\mycdot J_{A_n})\sum_{r}
{\frac{1}{\prod_{q\neq r}(2\epsilon\mycdot p_{1q})}}
\label{three_vertices}
\eea

Let us consider now a diagram that could occur from the ones above by contracting a propagator, thereby merging two of
the three-vertices, say the $r$'th and the $r+1$'th in one four-vertex. Such a diagram has one four-vertex on the line and one
propagator less, that is it has
$k$ vertices of which $k-1$ three-vertices and one four-vertex, as well as $k-1$ propagators.

\be
H(m_1,\ldots,\tilde{m}_r,\tilde{m}_{r+1},\ldots,m_{k+1};j)=
\begin{picture}(285,30) (-5,-10)
    \SetScale{0.4}
    \SetWidth{0.5}
    \SetColor{Black}
    \Line(15,-17)(75,-17)
    \Line(75,-17)(75,28)
    \Line(75,-17)(90,-17)
    \Line(135,-17)(165,-17)
    \Line(165,-17)(195,-17)
    \Line(165,-17)(150,28)
    \Line(165,-17)(180,28)
    \Line(240,-17)(300,-17)
    \Line(255,-17)(255,28)
    \GOval(146,41)(15,15)(0){0.411}
    \GOval(185,41)(15,15)(0){0.411}
    \GOval(75,41)(15,15)(0){0.411}
    \GOval(255,40)(15,15)(0){0.411}
    \Text(39,-7)[lb]{\small{\Black{$\ldots$}}}
    \Text(81,-7)[lb]{\small{\Black{$\ldots$}}}
    \Text(23,28)[lb]{\small{$m_1$}}
    \Text(46,28)[lb]{\small{$\tilde{m}_r$}}
    \Text(62,28)[lb]{\small{$\tilde{m}_{r+1}$}}
    \Text(95,28)[lb]{\small{$m_n$}}
    \Text(10,-18)[lb]{\small{$1$}}
    \Text(120,-18)[lb]{\small{$n$}}

  \end{picture}
\ee

The sum over all cuts of such hatted
diagrams is

\bea
\lefteqn{\sum_{\stackrel{j=1}{j\neq r}}^{k}{H_r(m_1,\ldots ,\tilde{m}_{r},\tilde{m}_{r},\ldots,m_k;j)}=}\nonumber\\& &=F(m_1,\ldots,\tilde{m}_{r},\tilde{m}_{r},\ldots, m_k)+
\nonumber\\
& &+H_{k,r}\sum_{\stackrel{j=1}{j\neq r}}^{k}{\left[\frac{z^{k-1}}{\prod_{q\neq j,r}\hat{p}_{1q}^2}\right]_{z=z_j}\frac{1}{p_{1j}^2}}
\eea
where $r$ denotes the position of the four-vertex. As before, all contributions involving $z^{\lambda}$ for $0<\lambda<k-1$ vanish due to
eq.\ref{vector_id}. Using the identity eq.\ref{full_id} we can perform the $j$ sum over cuts and get
\bea
\lefteqn{\sum_{\stackrel{j=1}{j\neq r}}^{k}{H_r(m_1,\ldots ,\tilde{m}_{r},\tilde{m}_{r},\ldots,m_k;j)}=}\nonumber\\& &=F(m_1,\ldots,\tilde{m}_{r},\tilde{m}_{r},\ldots, m_k)-
\tilde{H}_{k-1,r}\frac{1}{\prod_{q\neq r}{2\epsilon\mycdot p_{1q}}}
\eea
Here $\tilde{H}_{k-1,r}$ stands for the vertex contribution of a diagram with one four vertex and $k-1$ white-blob three-vertices. It is
not difficult to see that this is equal to
\be
\tilde{H}_{k-1,r}=(-1)^{k-1}(\epsilon_1\mycdot J_{A_1})(\epsilon\mycdot J_{A_2})\ldots(\epsilon_n \mycdot J_{A_{k}})
\ee
Summing up over all possible positions of the four vertex we get
\bea
\lefteqn{\sum_{r=1}^{k}\sum_{\stackrel{j=1}{j\neq r}}^{k}{H_r(m_1\ldots m_k;j)}=F(m_1,\ldots,\tilde{m}_{r},\tilde{m}_{r},\ldots, m_k)+
}\nonumber\\
& &+(-1)^{k}(\epsilon_1\mycdot J_{A_1})\ldots(\epsilon_n\mycdot J_{A_n})\sum_{r}{\frac{1}{\prod_{q\neq r}{2\epsilon\mycdot p_{1q}}}}
\eea
which exactly cancels the last term of equation eq.\ref{three_vertices}, thus completing the proof that the sum over all cuts of all possible
hatted diagrams is equal to the sum of Feynman diagrams.

\section{Remarks on generality}
We have seen in the previous sections how the BCFW decomposition is related to Feynman diagrams. In
fact, when working in the particular gauge that we have chosen, every Feynman diagram is broken in pieces
(`hatted' graphs) each of which contains a cut in one of the propagators along the line connecting the two `special' legs. Then,
hatted graphs with the same cut are regrouped together in one decomposition, and the sum of decompositions gives back the whole
color amplitude.

In the process of breaking up the Feynman graphs in hatted graphs, some extra terms are produced: those correspond
to the $z\rightarrow\infty$ limit of the hatted graph. Thanks to the particular structure of the YM vertices, demonstrated
in eq.\ref{vertices}, these extra terms cancel exactly among graphs with no four-vertices and graphs with one
four-vertex and one propagator less. In the lowest case of one propagator along the line, the diagram with two three vertices gives an extra term that is  exactly equal to
the diagram with no propagator and one four-vertex. The latter is a diagram which, having no propagator along the line,
didn't fit in an obvious way in any BCFW decomposition.

The absence of a Yang-Mills structure in a scalar theory is what makes a decomposition along the terms of BCFW cumbersome.
There one would have to accommodate the  diagram in eq.\ref{zerovertex1} by adding an ad hoc term in the recurrence relation
which would therefore be less elegant.

The gauge in which we are working explicitly eliminates the $z$-dependence from the polarization vectors.
In other gauges the polarization vectors $\epsilon_1^+$ and $\epsilon_n^-$ would be affected by the shift. This would complicate significantly the algebra, as further poles related to vanishing denominators
of these polarization vectors will come to play.

If the identity eq.\ref{zerovertex1} doesn't hold, for example, contributions from the class of
diagrams in the left hand side of eq.\ref{zerovertex1} would have to be cancelled by contributions from other classes. 
The appealing cancellations within subclasses of diagrams
with a particular partitioning of the external legs would be lost. Keeping track of the terms in fragmenting and regrouping Feynman
diagrams into BCFW decompositions would be much harder. Still one could prove the BCFW decomposition by performing the pole expansion
on any Feynman diagram minus its $z\rightarrow\infty$ limit but the algebra would be particularly cumbersome.

It is becoming increasingly clear that the BCFW decomposition is a rearrangement of terms of Feynman diagrams within a color ordered
gluonic amplitude. A similar impression will presumably occur from an approach of amplitudes with fermionic lines. 
Under the light of the diagrammatic proof presented here it is seen that the mechanism behind any apparent cancelations in the BCFW 
recursion formula does not (unfortunately) divulge a deeper principle or even a contingent effect in gluonic amplitudes, but results
directly from the structure of the Yang-Mills vertices and the cancellations induced by gauge invariance.
%From the point of view of computational complexity it was very early pointed out that the BCFW formula would be of little help. 
% Its merits as an analytic tool are not clear when compared with the Berends-Giele \cite{berends_giele}
%approach. 
\section{Acknowledgments}
Author P.D.D.  is currently co-funded by the European
Social Fund (75\%) and National Resources (25\%)-EPEAEK B! - PYTHAGORAS. The research of authors A.L. and C.G.P. has
been funded by the EU program no.HPMD-CT-2001-00105. The author C.G.P. 
also acknowledges HPC-EUROPA Programme for its support.

\end{document}